\newif\ifBASIC
\BASICfalse
\newif\ifWP
\WPfalse
\newif\ifJOURNAL
\JOURNALfalse

\newif\ifFULL
\FULLfalse
\newif\ifLATIN
\LATINfalse

\BASICtrue	% choose BASIC, WP (also used for arXiv), or JOURNAL
\LATINtrue	% LATIN means that the Cyrillic and Japanese references
\newif\ifnotJOURNAL	% derivative conditional
\notJOURNALtrue
\ifJOURNAL\notJOURNALfalse\fi

\newif\ifnotFULL	% derivative conditional
\notFULLtrue
\ifFULL\notFULLfalse\fi

\newif\ifnotLATIN	% derivative conditional
\notLATINtrue
\ifLATIN\notLATINfalse\fi

\ifBASIC
  \newcommand{\GTPxxviii}{GTP28arXiv}
\fi
\ifWP
  \newcommand{\GTPxxviii}{GTP28arXiv}
\fi
\ifJOURNAL
  \newcommand{\GTPxxviii}{vovk:FS}
\fi

\ifnotLATIN
  
\fi
\ifLATIN
  
\fi

\ifBASIC
  \documentclass[10pt]{article}
  \usepackage{amsmath,amsfonts,amssymb,amsthm,latexsym,graphicx}
\fi

\ifWP
  \documentclass[12pt]{article}
  \usepackage{amsmath,amsfonts,amssymb,amsthm,latexsym,graphicx}
  \input{/Doc/Computing/Latex/gt.txt}
\fi

\ifJOURNAL
  \documentclass[12pt,leqno]{article}
  \usepackage{amsmath,amsfonts,amssymb,amsthm,latexsym,graphicx}
\fi

\ifFULL
  \usepackage{color}

  \newcommand{\bluebegin}{\begingroup\color{blue}}
  \newcommand{\blueend}{\endgroup}

\fi

\allowdisplaybreaks[4]

\newcommand{\Vladimir}{Vladimir}

\ifnotLATIN
  \input{OT2enc.def}
  
  \usepackage{CJK}
\fi

\newcommand{\st}{\mathrel{|}}

\newcommand{\dd}{\mathrm{d}}	% differential (closes \int)

\newcommand{\K}{\mathcal{K}}	% capital (often called value here)

\DeclareMathOperator{\III}{\boldsymbol{1}}  % indicator
\DeclareMathOperator{\e}{\mathrm{e}}	    % e

\newcommand{\bbbr}{\mathbb{R}}			% the real numbers

\newcommand{\bbbe}{\mathbb{E}}		% auxiliary (expectation)
\DeclareMathOperator{\Expect}{\bbbe}

\newcommand{\bbbp}{\mathbb{P}}		% auxiliary (probability)

\DeclareMathOperator{\UpProb}{\lefteqn{\smash{\overline{\bbbp}}}\phantom{\bbbp}}
\DeclareMathOperator{\LowProb}{\underline{\bbbp}}	% lower probability

\theoremstyle{plain}
\newtheorem{theorem}{Theorem}[section]
\newtheorem{proposition}[theorem]{Proposition}
\newtheorem{corollary}[theorem]{Corollary}

\theoremstyle{definition}
\newtheorem*{remark}{Remark}

\makeatletter
  \@addtoreset{equation}{section}
  
\makeatother

\title{The efficient index hypothesis and its implications in the BSM model}
\ifBASIC
  \author{Vladimir Vovk}
  \date{September 11, 2011}
\fi

\ifWP
  \author{Vladimir Vovk}
\fi

\newcommand{\acknowledge}{The idea of this note
  originated in my attempts to understand Bodie's paradox \cite{bodie:1995},
  saying that it is expensive to insure against a shortfall of stock returns
  as compared to bond returns.
  I am grateful to Robert Merton for drawing my attention to Bodie's paper.
  Thanks to Wouter Koolen for illuminating discussions.
  The data for the empirical studies in Section~\ref{sec:EPP}
  have been provided by Yahoo!\ Finance and processed using R.
  This research has been supported in part
  by NWO Rubicon grant 680-50-1010.}

\begin{document}
\ifnotJOURNAL
  \maketitle
\fi

\ifJOURNAL
  \begin{titlepage}
  
  \vfill

  \begin{center}
    \makeatletter
    {\huge\@title}{\Large \footnotemark[1]}
    \makeatother
  \end{center}

  \medskip

  % \begin{center}
  % \iffalse
    \noindent
    \textbf{Vladimir Vovk} (Royal Holloway, University of London, UK)
    and \textbf{Wouter M. Koolen} (Royal Holloway, University of London, UK,
      and Centrum Wiskunde \&\ Informatica, Amsterdam, Netherlands)
  % \fi
  % \end{center}

  \bigskip

  \textbf{Key words:} index, efficient market hypothesis, equity premium puzzle

  \bigskip

  \begin{center}\textbf{\Large Abstract}\end{center}

\fi

\ifnotJOURNAL
  \begin{abstract}
\fi
  This note studies the behavior of an index $I_t$
  which is assumed to be a tradable security,
  to satisfy the BSM model $\dd I_t/I_t=\mu\dd t+\sigma\dd W_t$,
  and to be \emph{efficient} in the following sense:
  we do not expect a prespecified trading strategy
  whose value is almost surely always nonnegative
  to outperform the index greatly.
  The efficiency of the index imposes severe restrictions
  on its growth rate;
  in particular, for a long investment horizon
  we should have $\mu\approx r+\sigma^2$,
  where $r$ is the interest rate.
  This provides another partial solution to the equity premium puzzle.
  All our mathematical results are extremely simple.
  \ifFULL\bluebegin
    Our main technical tool is the substitution rule
    for evaluating integrals.
  \blueend\fi
\ifnotJOURNAL
  \end{abstract}
\fi

\ifJOURNAL

  \vfill

  \footnotetext[1]{\acknowledge}
  \end{titlepage}
\fi

\section{Introduction}

The \emph{efficient index hypothesis} (\emph{EIH})
is a version of the random walk hypothesis
and the efficient market hypothesis.
It is a statement about a specific index, such as S\&P 500,
and says that we do not expect a prespecified trading strategy
to beat the index by a factor of $1/\delta$ or more,
for a given threshold $\delta$ (such as $\delta=0.1$).
The trading strategy is assumed to be \emph{prudent},
in the sense of its value being nonnegative a.s.\ at all times.
% when started with initial capital 1
By saying that it beats the index by a factor of $1/\delta$ or more
we mean that its initial value is $\K_0>0$
and its final value $\K_T$ satisfies $\K_T/I_T\ge(1/\delta)(\K_0/I_0)$.
(By the value of a trading strategy
we always mean the undiscounted dollar value of its current portfolio.)
We will see that the EIH has several interesting implications,
such as $\mu\approx r+\sigma^2$ for the growth coefficient $\mu$ of the index.

We use the EIH in the interpretation of our results,
but their mathematical statements do not involve this hypothesis.
For example, in Section~\ref{sec:growth} we prove
that there is a prudent trading strategy that, almost surely,
beats the index by a factor of at least $10$ unless
\begin{equation}\label{eq:prop-2-sided-10}
  \frac{I_T}{\e^{rT}}
  \in
  \left(
    \e^{\sigma^2T/2-1.64\sigma\sqrt{T}},
    \e^{\sigma^2T/2+1.64\sigma\sqrt{T}}
  \right)
\end{equation}
(see Proposition~\ref{prop:2-sided}).
If we believe in the EIH (for $\delta=0.1$),
we should believe in (\ref{eq:prop-2-sided-10}).
But even if we do not believe in the EIH,
the proposition gives us a way of beating the index
when (\ref{eq:prop-2-sided-10}) is violated.

As used in this note,
the EIH is a weaker assumption than it appears to be.
There might be sophisticated prudent trading strategies that do beat the index
(by a large factor), but we are not interested in such strategies.
It is sufficient that the primitive strategies considered in this note
be not expected to beat the index.

Our EIH is obviously related,
and has a similar motivation, to the standard efficient markets hypothesis \cite{fama:1970}.
There are, however, important differences.
For example, the EIH does not assume that the security prices are ``correct'' in any sense,
or that investors' expectations are rational (individually or \emph{en masse}).
The EIH controls for risk only by insisting that our trading strategies
be prudent.
Admittedly, this is a weak requirement,
and so the threshold value of $\delta$ should be a small number;
in our examples, we use $\delta=0.1$.
(If a trader is worried about losing all money,
nothing prevents her from investing only part of her capital
in prudent strategies that can lose everything.)

\ifFULL\bluebegin
  The EIH is also related to the random walk hypothesis
  and to the ``martingale model''
  (refer to the Cootner symposium, Samuelson \cite{samuelson:1973},
  and Fama \cite{fama:1970}).
  It is very close to the hypothesis
  that, when discounted by dividing by $I_t$,
  the value of a prespecified prudent trading strategy is a supermartingale
  (w.r.\ to some ``reasonable'' probability measure,
  not w.r.\ to the ``physical'' measure that we postulate).
\blueend\fi

\begin{remark}
  In \cite{shafer/vovk:2001,vovk/shafer:2008CAPM},
  the EIH was referred to as the ``efficient market hypothesis'',
  whereas the standard hypothesis of market efficiency
  as the ``efficient markets hypothesis'',
  with ``markets'' in plural.
  However, nowadays the standard hypothesis
  is more often called the ``efficient market hypothesis''
  than the ``efficient markets hypothesis'',
  % (see, e.g., \cite{wikipedia:EMH})
  and so it is safer to use a different term for our hypothesis.
  The results of this note agree with the results of \cite{vovk/shafer:2008CAPM}
  (see, e.g., (1) of \cite{vovk/shafer:2008CAPM}
  as applied to $s_n:=r$, $\forall n$),
  which were obtained using very different methods.
\end{remark}

We start the main part of the note
with results about the growth rate of the index under the EIH
(Section~\ref{sec:growth}).
The main insight here is that the index outperforms the bond
approximately by a factor of $\e^{\sigma^2T/2}$
(cf.\ (\ref{eq:prop-2-sided-10})).
In the following section, Section~\ref{sec:mu},
we show that, under the EIH, $\mu\approx r+\sigma^2$.
Section~\ref{sec:EPP} applies this result to the equity premium puzzle;
the equity premium of $\sigma^2$ is closer to the observed levels
of the equity premium than the predictions of standard theories.
Section~\ref{sec:GTP} discusses our findings
from the point of view of game-theoretic probability
(see, e.g., \cite{shafer/vovk:2001}).
\ifFULL\bluebegin
  Finally, Section~\ref{sec:conclusion} states some directions of further research.
\blueend\fi

\ifFULL\bluebegin
  The index has to be defined as the average of \textbf{all} shares
  traded in the market.
  Something like idealized FTSE 100 or S\&P 500
  (containing the biggest 100 or 500 companies by market capitalization)
  will not work:
  such indexes cannot be used as tradable securities,
  because of the overheads (proportional to the local time)
  involved in replacing some of the 100 (or 500) companies
  by new companies (overtaking some old companies in the index).
\blueend\fi

\section{Growth rate of the index}
\label{sec:growth}

The time interval in this note is $[0,T]$, $T>0$;
in the interpretation of our results the horizon $T$
will be assumed to be a large number.
The value of the index at time $t$ is denoted $I_t$.
We assume that it satisfies the BSM (Black--Scholes--Merton) model
\begin{equation}\label{eq:physical}
  \frac{\dd I_t}{I_t}
  =
  \mu\dd t + \sigma\dd W_t
\end{equation}
and that $I_0=1$.
\ifFULL\bluebegin
  The value of the drift coefficient $\mu$ will matter little,
  and we can consider much more general drift terms using Girsanov's theorem
  without affecting our main conclusions.
\blueend\fi
The interest rate $r$ is assumed constant.
We will sometimes interpret $\e^{rt}$
as the price at time $t$ of a zero-coupon bond whose initial price is~$1$.

The risk-neutral version of (\ref{eq:physical}) is
\begin{equation}\label{eq:risk-neutral}
  \frac{\dd I_t}{I_t}
  =
  r\dd t + \sigma\dd W_t.
\end{equation}
The explicit strong solution to this SDE is
$
  I_t
  =
  \e^{(r-\sigma^2/2)t+\sigma W_t}
$.

Let $E\subseteq\bbbr$ be a Borel set.
If $F:\bbbr\to\bbbr$,
we let $F(E)$ stand for the set
$\{F(x)\st x\in E\}$;
this convention defines the meaning of expressions such as $\frac{\ln E}{2}-1$.
The BSM price at time $0$ of the European contingent claim
whose payoff at time $T$ is
\begin{equation}\label{eq:final}
  F(I_T)
  :=
  \begin{cases}
    I_T & \text{if $I_T\in E$}\\
    0 & \text{otherwise}
  \end{cases}
\end{equation}
can be computed as the discounted expected value
\begin{align}
  \e^{-rT}
  &\Expect
  \left(
    \e^{(r-\sigma^2/2)T+\sigma\sqrt{T}\xi}
    \III
    \left\{
      \e^{(r-\sigma^2/2)T+\sigma\sqrt{T}\xi}\in E
    \right\}
  \right)\notag\\
  &=
  \frac{1}{\sqrt{2\pi}}
  \int_{\frac{\ln E}{\sigma\sqrt{T}}-\frac{r}{\sigma}\sqrt{T}+\frac{\sigma}{2}\sqrt{T}}
  \e^{-\sigma^2 T/2 + \sigma\sqrt{T}x - x^2/2}
  \dd x\notag\\
  &=
  \frac{1}{\sqrt{2\pi}}
  \int_{\frac{\ln E}{\sigma\sqrt{T}}-\frac{r}{\sigma}\sqrt{T}+\frac{\sigma}{2}\sqrt{T}}
  \e^{-(x-\sigma\sqrt{T})^2/2}
  \dd x\notag\\
  &=
  \frac{1}{\sqrt{2\pi}}
  \int_{\frac{\ln E}{\sigma\sqrt{T}}-\frac{r}{\sigma}\sqrt{T}-\frac{\sigma}{2}\sqrt{T}}
  \e^{-y^2/2}
  \dd y\notag\\
  &=
  N_{0,1}
  \left(
    \frac{\ln E}{\sigma\sqrt{T}}
    -
    \frac{r}{\sigma}\sqrt{T}
    -
    \frac{\sigma}{2}\sqrt{T}
  \right),\label{eq:initial}
\end{align}
where $\xi\sim N_{0,1}$,
$N_{0,1}$ is the standard Gaussian distribution on $\bbbr$,
and $\III\{\ldots\}$ is defined to be $1$ if the condition in the curly braces
is satisfied and $0$ otherwise.

Since the BSM price can be hedged perfectly
(see, e.g., \cite{karatzas/shreve:1991}, Theorem 5.8.12),
% page 378
there is a prudent trading strategy $\Sigma$ with initial value (\ref{eq:initial})
and final value (\ref{eq:final}) a.s.
We can see that $\Sigma$ beats the market by the reciprocal to (\ref{eq:initial})
if $E$ happens.

\subsection*{Two special cases}

Let $\delta\in(0,1)$ and $E:=(-\infty,A]\cup[B,\infty)$,
where $A$ and $B$ are chosen such that
\begin{equation}\label{eq:A-B}
  \frac{\ln A}{\sigma\sqrt{T}}
  -
  \frac{r}{\sigma}\sqrt{T}
  -
  \frac{\sigma}{2}\sqrt{T}
  =
  -z_{\delta/2},
  \quad
  \frac{\ln B}{\sigma\sqrt{T}}
  -
  \frac{r}{\sigma}\sqrt{T}
  -
  \frac{\sigma}{2}\sqrt{T}
  =
  z_{\delta/2},
\end{equation}
where $z_p$ is the upper $p$-quantile of the standard Gaussian distribution,
i.e., is defined by the requirement that $N_{0,1}([z_{p},\infty))=p$.
Then $N_{0,1}(E)=\delta$.
Equations (\ref{eq:A-B}) give
$$
  A
  =
  \e^{rT}
  \e^{\sigma^2T/2-z_{\delta/2}\sigma\sqrt{T}},
  \quad
  B
  =
  \e^{rT}
  \e^{\sigma^2T/2+z_{\delta/2}\sigma\sqrt{T}}.
$$

We can state the result of our calculations as follows.
\begin{proposition}\label{prop:2-sided}
  Let $\delta>0$.
  There is a prudent trading strategy
  (depending on $\sigma,r,T,\delta$)
  that, almost surely, beats the index
  by a factor of $1/\delta$ unless
  \begin{equation}\label{eq:2-sided}
    \frac{I_T}{\e^{rT}}
    \in
    \left(
      \e^{\sigma^2T/2-z_{\delta/2}\sigma\sqrt{T}},
      \e^{\sigma^2T/2+z_{\delta/2}\sigma\sqrt{T}}
    \right).
  \end{equation}
\end{proposition}
\noindent
Equation~(\ref{eq:2-sided}) says that for large $T$ the efficient index
can be expected to outperform the bond $\e^{\sigma^2T/2}$-fold.
The case $\delta\ge1$ in Proposition \ref{prop:2-sided} is trivial,
but we do not exclude it to simplify the statement of the proposition.

If we are only interested in a lower or upper bound on $I_T$,
we should instead consider the set $E:=(-\infty,A]$ or $E:=[B,\infty)$,
respectively.
We will obtain (\ref{eq:A-B}) with $\delta$ in place of $\delta/2$
and, therefore, will obtain the following proposition.
\begin{proposition}\label{prop:1-sided}
  Let $\delta>0$.
  There is a prudent trading strategy that, almost surely, beats the index
  by a factor of $1/\delta$ unless
  \begin{equation*} % \label{eq:1-sided}
    \frac{I_T}{\e^{rT}}
    >
    \e^{\sigma^2T/2-z_{\delta}\sigma\sqrt{T}}.
  \end{equation*}
  There is another prudent trading strategy that, almost surely, beats the index
  by a factor of $1/\delta$ unless
  \begin{equation}\label{eq:1-sided}
    \frac{I_T}{\e^{rT}}
    <
    \e^{\sigma^2T/2+z_{\delta}\sigma\sqrt{T}}.
  \end{equation}
\end{proposition}

It is clear that Propositions~\ref{prop:2-sided} and~\ref{prop:1-sided}
are tight in the sense that the factor $1/\delta$ cannot be improved.

\section{Implications for $\mu$}
\label{sec:mu}

The following corollary of Proposition~\ref{prop:2-sided} shows
that the EIH and the BSM model (\ref{eq:physical})
imply $\mu\approx r+\sigma^2$.
% (we refrain from spelling out similar implications of Proposition~\ref{prop:1-sided})
\begin{proposition}\label{prop:mu}
  For each $\delta>0$ there exists a prudent trading strategy
  $\Sigma=\Sigma(\sigma,r,T,\delta)$
  that satisfies the following condition.
  % when started with initial capital 1
  For each $\epsilon>0$, either
  \begin{equation}\label{eq:mu}
    \left|
      r + \sigma^2 - \mu
    \right|
    <
    \frac{(z_{\delta/2}+z_{\epsilon})\sigma}{\sqrt{T}}
  \end{equation}
  or $\Sigma$ beats the index by a factor of at least $1/\delta$
  with probability at least $1-\epsilon$.
\end{proposition}
\noindent
Intuitively, $\mu\approx r+\sigma^2$
unless we can beat the index or a rare event happens
(assuming that $\delta$ and $\epsilon$ are small and $T$ is large).
\begin{proof}[Proof of Proposition~\ref{prop:mu}]
  Without loss of generality, assume $\delta,\epsilon\in(0,1)$.
  As $\Sigma$ we take a prudent trading strategy
  that beats the index by a factor of $1/\delta$ unless (\ref{eq:2-sided}) holds.
  Therefore, we are only required to prove
  that the event that (\ref{eq:2-sided}) holds but (\ref{eq:mu}) does not
  has probability at most $\epsilon$.
  We can rewrite (\ref{eq:2-sided}) as
  \begin{equation}\label{eq:step-1}
    \left|
      \ln I_T - rT - \frac{\sigma^2}{2}T
    \right|
    <
    z_{\delta/2}\sigma\sqrt{T}.
  \end{equation}
  Remembering that (\ref{eq:physical}) has explicit solution
  $
    I_t
    =
    \e^{(\mu-\sigma^2/2)t+\sigma W_t}
  $,
  we can rewrite (\ref{eq:step-1}) as
  \begin{equation*} % \label{eq:step-2}
    \left|
      \sigma\sqrt{T}\xi
      -
      (r+\sigma^2-\mu)T
    \right|
    <
    z_{\delta/2}\sigma\sqrt{T},
  \end{equation*}
  where $\xi\sim N_{0,1}$,
  i.e., as
  \begin{equation}\label{eq:step-3}
    \left|
      \xi
      -
      \frac{(r+\sigma^2-\mu)\sqrt{T}}{\sigma}
    \right|
    <
    z_{\delta/2}.
  \end{equation}
  If (\ref{eq:mu}) is violated, we have
  either $r+\sigma^2-\mu<-(z_{\delta/2}+z_{\epsilon})\sigma/\sqrt{T}$
  or $r+\sigma^2-\mu>(z_{\delta/2}+z_{\epsilon})\sigma/\sqrt{T}$.
  The two cases are analogous,
  and we consider only the first.
  In this case, (\ref{eq:step-3}) implies $\xi<-z_{\epsilon}$,
  the probability of which is $\epsilon$.
\end{proof}

\ifFULL\bluebegin
  Should I specialize Proposition~\ref{prop:mu} to $\epsilon=\delta$?
  This would be awkward since I take $\delta:=0.1$
  and the standard values for $\epsilon$ are $0.05$ and $0.01$.
\blueend\fi

Proposition~\ref{prop:mu} shows that the arbitrariness of $\mu$
in the BSM model (\ref{eq:physical}) for the index is to a large degree illusory
if we accept the EIH.

The strategy $\Sigma$ of Proposition~\ref{prop:mu}
depends only on $\sigma$, $r$, $T$, and $\delta$.
If we allow, additionally, dependence on $\mu$ and $\epsilon$,
we can use Proposition~\ref{prop:1-sided} instead of Proposition~\ref{prop:2-sided}
and strengthen (\ref{eq:mu}) by replacing $\delta/2$ with $\delta$.
\begin{proposition} % \label{prop:mu-bis}
  Let $\delta>0$ and $\epsilon>0$.
  Unless
  \begin{equation}\label{eq:mu-bis}
    \left|
      r + \sigma^2 - \mu
    \right|
    <
    \frac{(z_{\delta}+z_{\epsilon})\sigma}{\sqrt{T}},
  \end{equation}
  there exists a prudent trading strategy
  $\Sigma=\Sigma(\mu,\sigma,r,T,\delta,\epsilon)$
  that beats the index by a factor of at least $1/\delta$
  with probability at least $1-\epsilon$.
\end{proposition}
\begin{proof}
  Suppose~(\ref{eq:mu-bis}) is violated.
  Since the cases $r+\sigma^2-\mu<-(z_{\delta}+z_{\epsilon})\sigma/\sqrt{T}$
  and $r+\sigma^2-\mu>(z_{\delta}+z_{\epsilon})\sigma/\sqrt{T}$
  are analogous,
  we will assume
  \begin{equation}\label{eq:case}
    r+\sigma^2-\mu
    <
    -\frac{(z_{\delta}+z_{\epsilon})\sigma}{\sqrt{T}}.
  \end{equation}
  (Our trading strategy depends on which of the two cases holds,
  and so depends on $\mu$ and $\epsilon$.)
  As $\Sigma$ we take a prudent trading strategy
  that beats the index by a factor of $1/\delta$ unless (\ref{eq:1-sided}) holds.
  We are required to prove
  that the probability of (\ref{eq:1-sided}) is at most $\epsilon$.
  We can rewrite (\ref{eq:1-sided}) as
  \begin{equation*} % \label{eq:step-1-bis}
    \ln I_T - rT - \frac{\sigma^2}{2}T
    <
    z_{\delta}\sigma\sqrt{T},
  \end{equation*}
  i.e.,
  \begin{equation*} % \label{eq:step-3-bis}
    \xi
    -
    \frac{(r+\sigma^2-\mu)\sqrt{T}}{\sigma}
    <
    z_{\delta},
  \end{equation*}
  where $\xi\sim N_{0,1}$.
  The last inequality and (\ref{eq:case}) imply $\xi<-z_{\epsilon}$,
  whose probability is $\epsilon$.
\end{proof}

\ifFULL\bluebegin
  For the infinite horizon, we can prove:
  \begin{corollary}
    $\mu=r+\sigma^2$ unless we can beat the index almost surely.
  \end{corollary}
\blueend\fi

\section{Equity premium puzzle}
\label{sec:EPP}

The equity premium is the excess of stock returns over bond returns,
and it appears to be higher in the real world
than suggested by standard economic theories.
This has been dubbed the equity premium puzzle
% by Rajnish Mehra and Edward Prescott in their 1985 paper
\cite{mehra/prescott:1985}.
There is no consensus as to the explanation, or even to the existence,
of the equity premium puzzle;
for recent reviews see, e.g., \cite{mehra:2006,mehra/prescott:2008}.
\ifFULL\bluebegin
  See also
  \cite{kocherlakota:1996,siegel/thaler:1997,mehra:2003,mehra:2006}.
\blueend\fi
In this section we will see that our results can be interpreted
as providing a partial solution to the puzzle.

According to Proposition~\ref{prop:mu},
under the EIH we can expect $\mu\approx r+\sigma^2$.
This gives the equity premium $\sigma^2$.
The annual volatility of S\&P 500 is approximately 20\%
(see, e.g., \cite{mehra/prescott:2008chap1}, p.~3,
or \cite{mehra:2006}, p.~8),
which translates into an expected 4\% equity premium.
The standard theory predicts an equity premium of at most 1\%
(\cite{mehra:2006}, p.~11).

The empirical study by Mehra and Prescott
reported in \cite{mehra/prescott:2008chap1}, Table~2,
estimates the equity premium over the period 1889--2005 as $6.36\%$.
Taking into account the later years 2006--2010
reduces it, but not much, to $6.05\%$.
(The recent news about bonds outperforming stocks over the past 30 years
\cite{bloomberg:2009}
were about 30-year Treasury bonds,
whereas Mehra and Prescott use short-term Treasury bills for this period.)
Our figure of $4\%$ is below $6.05\%$,
but the difference is much less significant
than for the standard theory.
If the years 1802--1888 are also taken into account
(as done by Siegel \cite{siegel:1992},
updated until 2004 by Mehra and Prescott \cite{mehra/prescott:2008chap1}, Table~2,
and until 2010 by myself),
the equity premium goes down to $5.17\%$.

Equation~(\ref{eq:2-sided}) allows us to estimate the accuracy
of our estimate $\sigma^2$ of the equity premium.
Namely, we have, almost surely,
\begin{equation}\label{eq:accuracy}
  \frac1T
  \int_0^T
  \frac{\dd I_t}{I_t}
  -
  r - \sigma^2
  =
  \frac{\ln I_T+\sigma^2T/2-rT-\sigma^2T}{T}
  \in
  \left(
    -\frac{z_{\delta/2}\sigma}{\sqrt{T}},
    \frac{z_{\delta/2}\sigma}{\sqrt{T}}
  \right)
\end{equation}
unless a prespecified prudent trading strategy beats the index
by a factor of $1/\delta$.
Plugging $\delta:=0.1$ (to obtain a reasonable accuracy),
$\sigma:=0.2$, and $T:=2010-1888$,
we evaluate $z_{\delta/2}\sigma/\sqrt{T}$ in (\ref{eq:accuracy}) to $2.98\%$
for the period 1889--2010,
and changing $T$ to $2010-1801$,
we evaluate it to $2.28\%$ for the period 1802--2010.
For both periods,
the observed equity premium falls well within the prediction interval.

\section{Three kinds of probabilities for the index}
\label{sec:GTP}

In this section we will take a broader view
of the simple results of the previous sections.
We started from the ``physical'' probability measure (\ref{eq:physical}),
used the risk-neutral probability measure (\ref{eq:risk-neutral}),
and saw the importance of the ``EIH measure''
\begin{equation}\label{eq:EIH}
  \frac{\dd I_t}{I_t}
  =
  (r+\sigma^2) \dd t
  +
  \sigma \dd W_t.
\end{equation}
We will see that the last two are essentially special cases of game-theoretic probability,
as defined in \cite{shafer/vovk:2001}.
If $E$ is a Borel subset of the Banach space $\Omega:=C([0,T])$
of all continuous functions on $[0,T]$,
we define its \emph{upper probability with bond as num\'eraire} by
$$
  \UpProb_{{\rm b}}(E)
  :=
  \inf
  \left\{
    \K_0
    \left|\;
      \frac{\K_T}{\e^{rT}} \ge \III_E \text{ a.s.}
    \right.
  \right\},
$$
where $\III_E$ is the indicator function of $E$,
$\K$ ranges over the value processes of prudent trading strategies,
and ``a.s.''\ means with probability one
under the physical measure (\ref{eq:physical})
(equivalently, under (\ref{eq:risk-neutral}) or under (\ref{eq:EIH})).
In other words,
$\UpProb_{{\rm b}}(E)$ is the infimum of $\delta>0$
such that a prudent trading strategy can beat the bond
by a factor of $1/\delta$ or more on the event $E$
(except for its subset of zero probability).
We define the \emph{upper probability} of $E$ \emph{with index as num\'eraire} by
$$
  \UpProb_I(E)
  :=
  \inf
  \left\{
    \K_0
    \left|\;
      \frac{\K_T}{I_T} \ge \III_E \text{ a.s.}
    \right.
  \right\}.
$$
In other words,
$\UpProb_I(E)$ is the infimum of $\delta>0$
such that a prudent trading strategy can beat the index
by a factor of $1/\delta$ or more on the event $E$.

For each Borel $E$,
$\UpProb_{{\rm b}}(E)$ is the risk-neutral measure of $E$
and $\UpProb_I(E)$ is its EIH measure.
\ifFULL\bluebegin
  Let us check the last statement for events of the form $I_T\in E$.
  The explicit solution of (\ref{eq:EIH}) is
  $\e^{(r+\sigma^2/2)t+\sigma W_t}$.
  Solving $\e^{(r+\sigma^2/2)t+\sigma\sqrt{T}\xi}\in E$ in $\xi$,
  we obtain $\xi\in A$,
  where $A$ is the expression in the parentheses in (\ref{eq:initial}).
\blueend\fi
It is standard in game-theoretic probability to define
the corresponding \emph{lower probabilities}
$$
  \LowProb_{{\rm b}}(E)
  :=
  1 - \UpProb_{{\rm b}}(E^c)
  \text{ and }
  \LowProb_I(E)
  :=
  1 - \UpProb_I(E^c),
$$
where $E^c:=\Omega\setminus E$.
Since our market is complete,
upper and lower probabilities always coincide.
A major difference of the definitions of $\UpProb_{{\rm b}}$ and $\UpProb_I$
from the usual definitions of upper probabilities
in game-theoretic probability is the presence of ``a.s.'';
in game-theoretic probability ``a.s.'' is absent
as there is no probability measure to begin with.

The processes (\ref{eq:risk-neutral}) and (\ref{eq:EIH})
are in some sense reciprocal.
By It\^o's formula,
if $I_t$ satisfies (\ref{eq:risk-neutral}),
then $I^*_t:=\e^{2rt}/I_t$ will satisfy (\ref{eq:EIH})
with $I^*$ in place of $I$ and $-W$ in place of $W$,
and vice versa.
(The definition of $I^*_t$ makes the bond's growth rate $\e^{rt}$
the geometric mean of $I_t$ and $I^*_t$.)
\ifFULL\bluebegin
  This may be another manifestation of the duality between the bond and the index,
  as discussed,
  in the case of discrete time and zero interest rate,
  in \cite{koolen/derooij:2010}, Section 2.3.
\blueend\fi
In particular,
the growth rate of typical trajectories of (\ref{eq:risk-neutral})
is approximately $\e^{(r-\sigma^2/2)t}$,
and the growth rate of typical trajectories of (\ref{eq:EIH})
is approximately $\e^{(r+\sigma^2/2)t}$.

\ifFULL\bluebegin
  These are the details of the application of It\^o's formula.
  Define $F(x,t):=\e^{2rt}x^{-1}$;
  then $F_x(x,t)=-\e^{2rt}x^{-2}$,
  $F_{xx}(x,t)=2\e^{2rt}x^{-3}$,
  and $F_t(x,t)=2r\e^{2rt}x^{-1}$.
  As
  $$
    \dd I_t = I_t r \dd t + I_t \sigma \dd W_t,
  $$
  we have
  \begin{align*}
    \dd I^*_t
    &=
    \dd F(I_t,t)
    =
    F_x(I_t) \dd I_t
    +
    \frac12 F_{xx}(I_t) I_t^2 \sigma^2 \dd t
    +
    F_t(I_t) \dd t\\
    &=
    -\e^{2rt}
    I_t^{-2}
    (I_t r \dd t + I_t \sigma \dd W_t)
    +
    \e^{2rt} I_t^{-3} I_t^2 \sigma^2 \dd t
    +
    2r \e^{rt} I_t^{-1} \dd t\\
    &=
    - I_t^* r \dd t
    -
    I^*_t \sigma \dd W_t
    +
    I^*_t \sigma^2 \dd t
    +
    2r I_t^* \dd t
    =
    I^*_t (r+\sigma^2) \dd t
    -
    I^*_t \sigma \dd W_t.
  \end{align*}
\blueend\fi

\ifFULL\bluebegin
  \section{Conclusion}
  \label{sec:conclusion}

  Directions of further research:
  \begin{itemize}
  \item
    Extend the results of this note to the probability-free setting
    of \cite{\GTPxxviii}.
  \item
    Can we develop the analogue of the game-theoretic CAPM of \cite{vovk/shafer:2008CAPM}
    in the current framework
    (with the number independent BMs greater than the number of assets,
    to make the market incomplete).
  \item
    If $I_t$ is a stock price and it underperforms the $\sigma^2\dd t$ drift,
    how rich can we become?
    (Is there a hope of reaching, or approaching, the $\sigma^2\dd t$ level?)
    The drift coefficient can be improved: $\mu\mapsto F(\mu)$ unless $\mu=\sigma^2$;
    what is $F$?
    We should not have a hard $\delta$.
  \end{itemize}

  \appendix
  \section{Rigorous sources for continuous-time mathematical finance}

  \begin{itemize}
  \item
    Shiryaev \cite{shiryaev:1999};
    perfect hedging is the subject of Section VII.4b:
    see Theorem~2 on p.~710.
    How does this theorem take the interest rate into account?
  \item
    Karatzas and Shreve's textbook \cite{karatzas/shreve:1991}
    gives an excellent description, both rigorous and intuitive:
    see Theorem 5.8.12 on p.~378.
  \end{itemize}
\blueend\fi

\ifnotJOURNAL
\section*{Acknowledgments}

\acknowledge
\fi

\end{document}